\documentclass[cits]{PoS}

\pdfoutput=1

\usepackage{verbatim,dsfont}

\def\tr{\mbox{tr }}

\def\ts{\tilde{\Sigma}}

\title{Dressed Polyakov loops and center symmetry from Dirac spectra}

\ShortTitle{Dressed Polyakov loops ...}

\author{\speaker{Falk Bruckmann} and Christian Hagen\\
        Institute for Theoretical Physics, University of Regensburg, 
        D-93040 Regensburg, Germany}

\author{Erek Bilgici and Christof Gattringer\\
        Institut f\"ur Physik, FB Theoretische Physik, Universit\"at Graz,
        A-8010 Graz, Austria\\ \\
        E-mails:
        \email{erek.bilgici@uni-graz.at, falk.bruckmann@physik.uni-r.de,
        christof.gattringer@uni-graz.at, christian.hagen@physik.uni-r.de}
        }

\abstract{
We construct a novel observable for finite temperature QCD that relates confinement and chiral symmetry.
It uses phases as boundary conditions for the fermions. 
We discuss numerical and analytical aspects of this observable,
like its spectral behavior below and above the critical temperature,
as well as the connection to chiral condensate and
center symmetry.
}

\FullConference{8th Conference Quark Confinement and the Hadron Spectrum \\
                 September 1-6 2008\\
                 Mainz, Germany}

\begin{document}


\noindent The main phenomena in QCD at the finite temperature transition
are deconfinement and chiral symmetry restoration.
There has always been the question whether there is a mechanism connec\-ting the two,
in particular since in the quenched theory the corresponding phase transitions
occur at the same critical temperatures \cite{kogut:83}.
Here we define an observable that indeed links the two
and discuss numerical findings from quenched lattice configurations
\cite{bilgici:08}.

The Polyakov loop, on the one hand,
is the order parameter of confinement, being traceless in the confined phase and
moving towards the center of the gauge group 
at temperatures above $T_c$.
The spectral density of the Dirac operator at the origin, $\rho(0)$, 
on the other hand,
is the order parameter of chiral symmetry. 
It is related to the condensate by the famous Banks-Casher relation~\cite{banks:80}
$\langle\bar{\psi}\psi\rangle=-\pi\rho(0)$ and vanishes above $T_c$.
How does confinement leave a trace in the Dirac spectrum? 
The answer lies in the dependence on temporal boundary conditions, as we will show now.


We use the lattice as a regulator.
The (untraced) Polyakov loop
is the product of temporal links.
For the lattice Dirac operator we use the staggered one \cite{kogut:74a}, 
which can be viewed as hopping by one link.
It is 
well-known that the $k$-th power of the Dirac operator at the same argument,
$D^k(x,x)$, contains all products of links along closed loops of length $k$ 
starting 
at $x$.
To distinguish the Polyakov loop from `trivially closed' loops 
(like 
the plaquette), 
%
one needs phase boundary conditions 
for the fermions~\cite{gattringer:06b},
$\psi(x_0+\beta,\vec{x})=e^{i\varphi} \psi(x_0,\vec{x})$.
These boundary conditions
amount to an imaginary chemical potential. 
They can be easily implemented by replacing the temporal links $U_0$
in some time slice by $e^{i\varphi}U_0$. 
As a consequence, all 
loops get a factor $e^{i\varphi q}$ 
according to their winding number $q$ around the temporal direction,
trivially closed loops remain unchanged.

In this way, the Polyakov loop can be reconstructed from the Dirac spectrum 
by using at least three boundary conditions, see \cite{bruckmann:06b}.
This `thin' Polyakov loop, however, has poor renormalization and scaling properties
and it turned out that in this approach it is UV dominated \cite{bruckmann:06b}.

Influenced by the Jena group \cite{synatschke:07a}, we instead consider the propagator $1/(m+D)$ with some probe mass $m$
and use a geometric series obtaining all powers of the Dirac operator.
Denoting the Dirac operator at a particular boundary condition $\varphi$ by $D_\varphi$, i.e., including the factors of $e^{i\varphi}$,
the propagator is given as a product of links along all closed loops $l$,
\begin{equation}
 \tr\,\frac{1}{m+D_{\varphi}}=
  \frac{1}{m}\sum_{{\rm loops}\:\: l}
  \frac{{\rm sign}(l)}{(2am)^{|l|}}\:
e^{\,i\varphi q(l)}\:\,
\tr_{\!c}\!\!\!\!\prod_{(x,\mu)\in l} U_\mu(x)\,,
\label{eq_pre_formula}
\end{equation}
where $|l|$ is the length of the loop and ${\rm sign}(l)$ comes from the staggered factor.

Of importance in (\ref{eq_pre_formula}) is the phase factor, 
since one can project onto a particular winding $q$ by a Fourier transform w.r.t. the boundary angle,
$\int_0^{2\pi} d\varphi \, e^{-i\varphi q}\ldots$.
Specifying to a single winding, $q=1$, like for the Polyakov loop, we arrive at \cite{bilgici:08}
\begin{equation}
 \ts\equiv
 \int_0^{2\pi} \frac{d\varphi}{2\pi} \, e^{-i\varphi}
 \frac{1}{V}\Big\langle \tr \frac{1}{m+D_\varphi}
 \Big\rangle = 
 \frac{1}{mV}
\sum_{q(l)=1}
 \frac{{\rm sign}(l)}{(2am)^{|l|}}\:
\Big\langle\tr_{\!c}\!\!\!\!\prod_{(x,\mu)\in l} U_\mu(x)\Big\rangle\,.
\label{eq_the_formula}
\end{equation}
We refer to the new observable $\ts$ as the `dual condensate',
because it is obtained through a Fourier transform
from the trace of the propagator.
Indeed, in the massless limit (after the infinite volume limit as usual) 
we obtain the chiral condensate 
integrated with a phase factor over the boundary conditions
and, using the Banks-Casher relation at every individual angle $\varphi$,
the corresponding representation in terms of the eigenvalue densities 
$\rho(0)_\varphi$. 
The right hand side of (\ref{eq_the_formula}) represents the `dressed Polyakov loop', 
that is the set of all loops
which wind once around the temporal direction. 
In the infinite mass limit, detours become suppressed and only the straight 
Polyakov loop survives as it is the shortest possible loop in this set.


\begin{figure}[t]
 \includegraphics[width=0.46\linewidth]{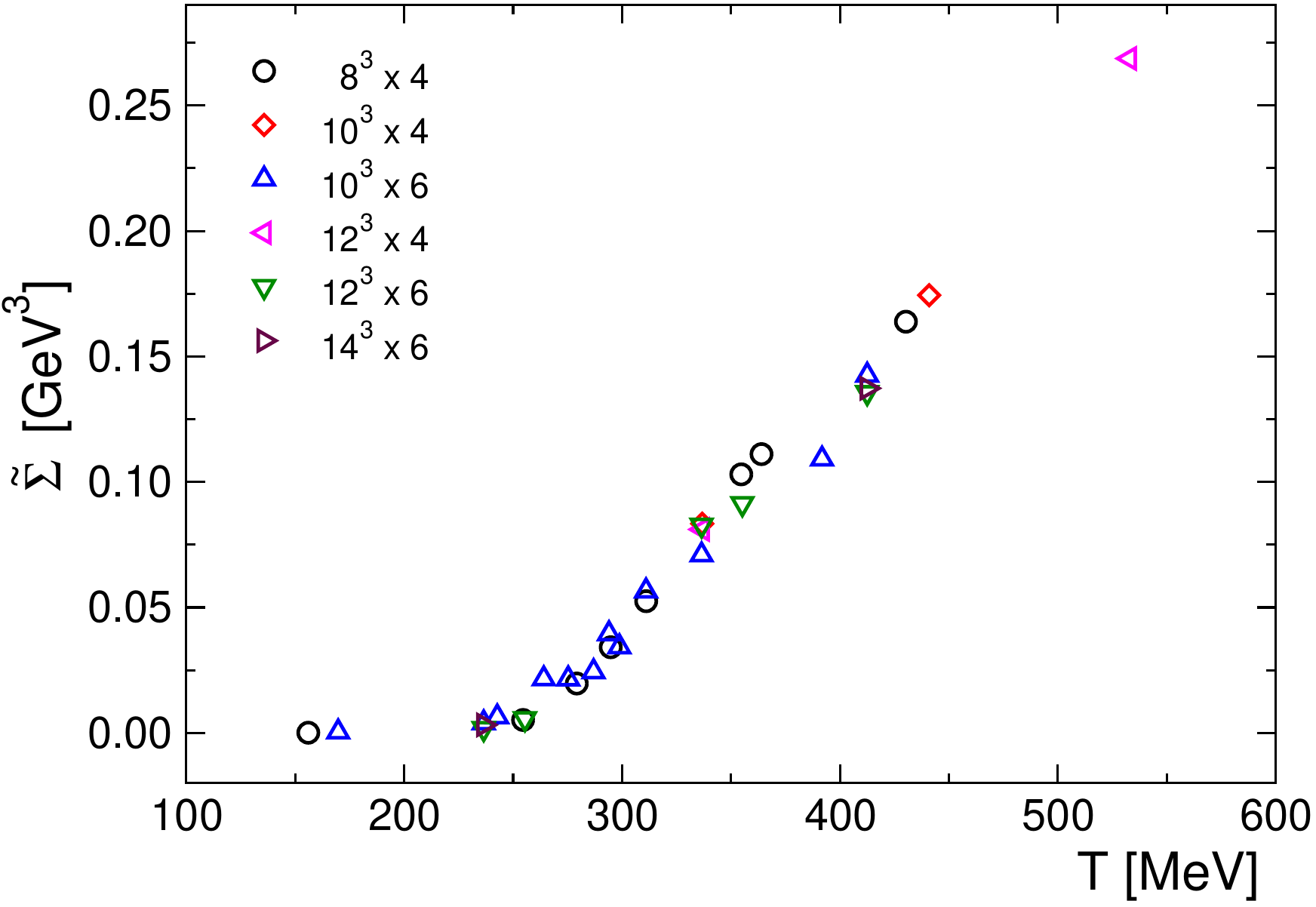}\hfill
 \includegraphics[width=0.46\linewidth]{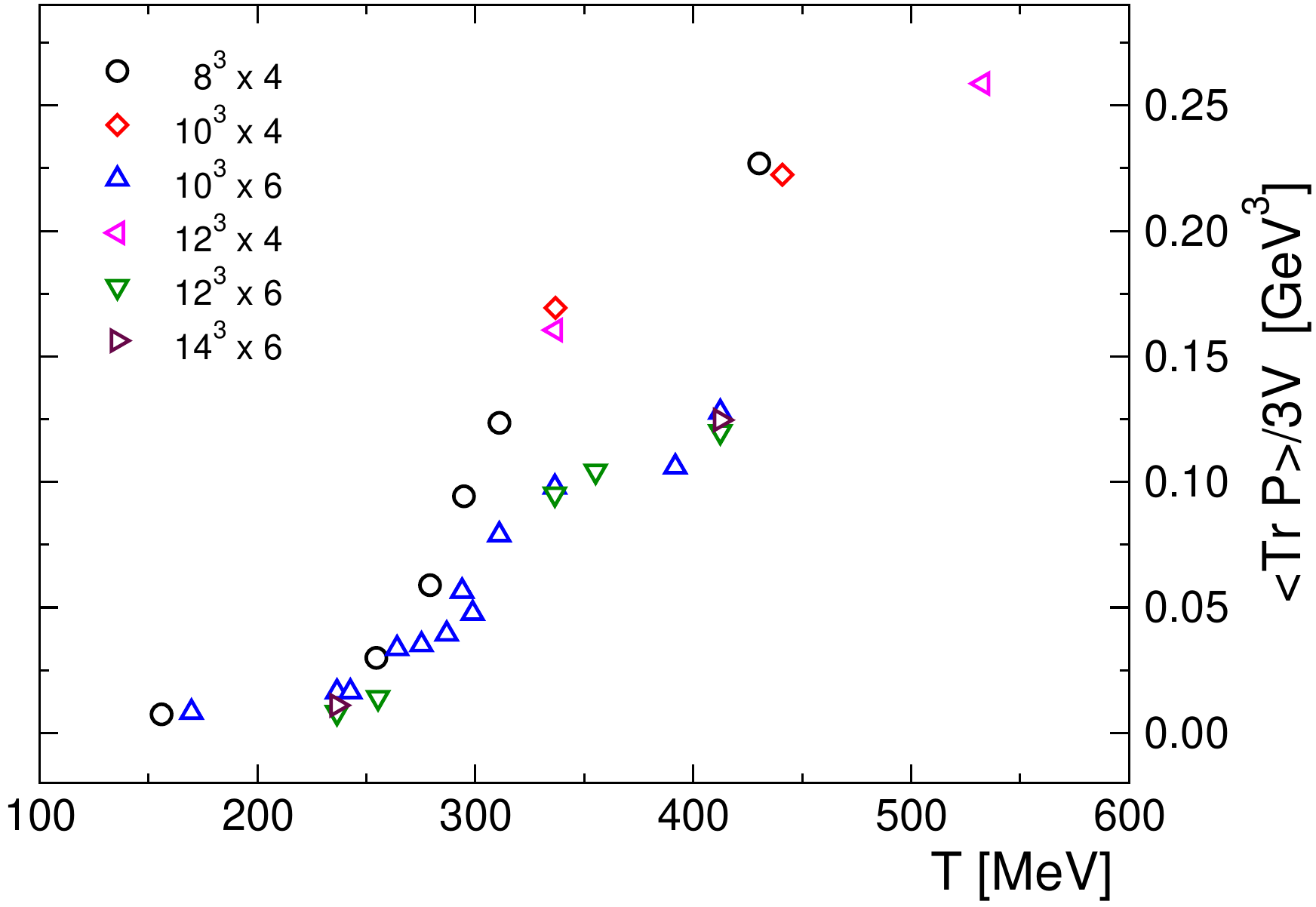}
 \caption{Left: The dressed Polyakov loop $\ts$ as a function of $T$ 
at various lattices (for $m =$ 100 MeV).
Right: The corresponding plot for the conventional Polyakov loop.}
\label{fig_order_parameter}
\end{figure}


Fig.~\ref{fig_order_parameter} shows
that $\ts$ is indeed an order parameter. 
Keeping the mass $m$ fixed, $\ts$ vanishes 
below the critical temperature 
(which is about 280 MeV in the quenched case)
and develops an expectation value for higher temperatures.
One finds that the results (when expressed in physical units)
obtained for different volumes 
and with different resolution 
essentially fall on a universal curve.
This illustrates the good renormalization properties of our observable,
which 
are inherited from $\langle\bar{\psi}\psi\rangle$
and may be viewed as an effect of
the dressing rendering such loops less UV dominated.

\begin{figure}[!b]
 \begin{center}
 \includegraphics[width=0.6\linewidth]{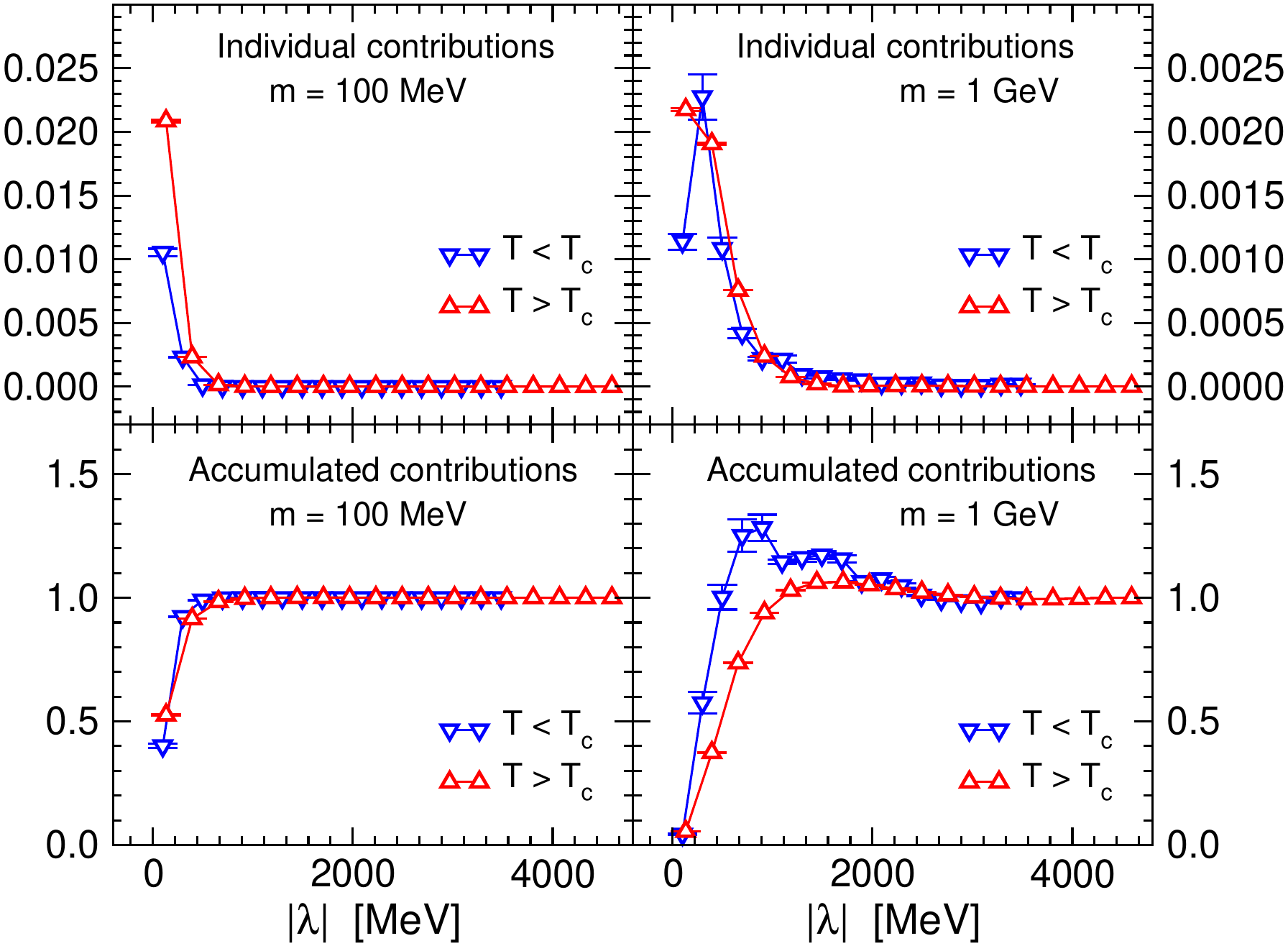}
 \end{center}
\vspace{-0.6cm}
 \caption{Individual and accumulated contributions 
 to the spectral sum (\protect\ref{eq_spectral_rep}) at two different masses.}
 \label{fig_spectral_contributions}
\end{figure}

In a spectral representation, the Dirac operator in 
$\ts$ is replaced by an eigenvalue sum, 
\begin{equation}
\ts=\int_0^{2\pi} \frac{d\phi}{2\pi} \, e^{-i\phi}
 \frac{1}{V}\Big\langle \sum_i 
\frac{1}{m+\lambda_\phi^{(i)}}\Big\rangle\,.
\label{eq_spectral_rep}
\end{equation}
As the eigenvalues appear in the denominator, 
we expect the sum to be dominated by the IR modes.
As Fig.~\ref{fig_spectral_contributions} shows, this is confirmed by our lattice data.

\begin{figure}[!t]
 \begin{center}
 \includegraphics[width=0.45\linewidth]{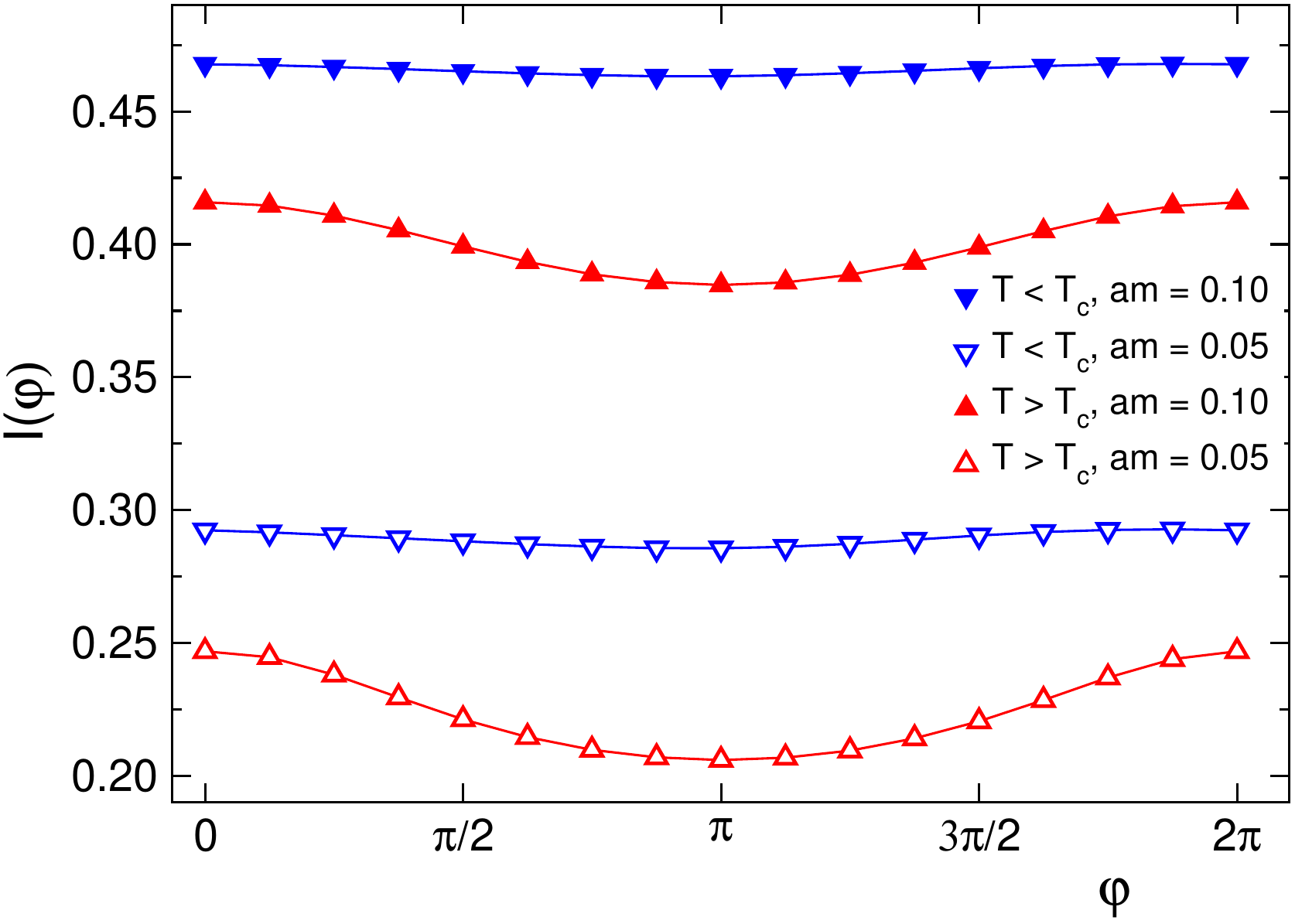}
 \end{center}
\vspace{-0.6cm}
 \caption{The integrand $\langle \sum_i (m+\lambda_\varphi^{(i)})^{-1}\rangle/V$
as a function of the boundary condition $\varphi$ for two values of $am$ 
and two temperatures,
 for configurations with real Polyakov loop.
}
\label{fig_response_integrand}
\end{figure}

How is a finite resp.\ vanishing order parameter $\ts$ 
built up by the eigenvalues? 
Fig.~\ref{fig_response_integrand} shows that they respond differently to the
boundary conditions in the confined vs.\ deconfined phase.
The eigenvalues are independent of the boundary condition in the confined phase, 
which leads to a vanishing order parameter $\ts$. 
In the deconfined phase, on the other hand, 
the eigenvalues show a typical cosine-type of modulation.
Together with the Fourier factor this yields a nonvanishing $\ts$. 

The chiral condensate has to behave essentially in the same way, 
as it is the integrand in the massless limit.
Although the chiral condensate is finite in the confined phase, 
it is independent of the boundary condition $\varphi$ 
and hence results in a vanishing dual condensate. 
The situation in the deconfined phase might be more confusing at first glance
as the spectrum has a gap there.
For boundary conditions in line with the original Polyakov loop,
however,
the chiral condensate persists above the critical temperature $T_c$, which has been seen in several studies \cite{gattringer:02b}.
This mechanism is needed to ensure a finite $\ts$ and should actually be at work for all $T>T_c$.

In the quenched case the conventional Polyakov loop is the order parameter for center symmetry.
Under center transformations the dressed Polyakov loop behaves in the same way.
Hence, $\ts$ is an order parameter for center symmetry, which is underneath our numerical findings and 
which we therefore expect to be independent of the choice of the lattice Dirac operator.

\end{document}